\documentclass[a4paper, 14pt, conference,onecolumn]{ieeeconf}      % Use this line for a4 paper

\IEEEoverridecommandlockouts                              % This command is only needed if 
\usepackage{graphicx}      % include this line if your document contains figures
\usepackage{amsmath}
\usepackage{fancyhdr}
\usepackage[english]{babel}
\usepackage[latin1]{inputenc}   % Neccessaire pour paragraphe (Harald)
\usepackage{amsfonts,amsmath,bm,hhline}
\usepackage{epsfig}
\usepackage{graphicx}
\usepackage{color}
\usepackage[hang]{caption2}
\usepackage[normalsize,it,hang]{subfigure}
\newtheorem{remark}{Remark}  
\title{
{\normalsize MATHMOD 2022 \\ 10th Vienna International Conference on Mathematical Modelling \\
\vspace{0.1cm}
July 27 - 29, 2022  -- Vienna, Austria} \\
\vspace{1.5cm}
{\Huge \bf Feedback control of social distancing for COVID-19 via elementary formulae}
}

\author{Michel Fliess$^{1,4}$, C\'edric Join$^{2,4}$ and Alberto d'Onofrio$^{3}$% <-this % stops a space
\thanks{$^{1}${LIX (CNRS, UMR 7161), \'Ecole polytechnique, 91128 Palaiseau, France \newline  {\tt Michel.Fliess@polytechnique.edu}}}
\thanks{$^{2}${CRAN (CNRS, UMR 7039)), Universit\'{e} de Lorraine, BP 239, 54506 Vand{\oe}uvre-l\`{e}s-Nancy, France \newline {\tt cedric.join@univ-lorraine.fr}}}
\thanks{$^{3}${Institut Camille Jordan, Universit\'{e} Claude Bernard Lyon 1, 69622 Villeurbanne, France \newline {\tt adonofrio1967@gmail.com}}}
\thanks{$^{4}${AL.I.E.N., 7 rue Maurice Barr\`{e}s, 54330 V\'{e}zelise, France \newline {\tt \{michel.fliess, cedric.join\}@alien-sas.com}}}
}

\begin{document}

\maketitle
\thispagestyle{empty}
\pagestyle{empty}

%%%%%%%%%%%%%%%%%%%%%%%%%%%%%%%%%%%%%%%%%%%%%%%%%%%%%%%%%%%%%%%%%%%%%%%%%%%%%%%%
\begin{abstract}
%This paper is concerned with the security of power network against external attacks. 
{\normalsize Social distancing has been enacted in order to mitigate the spread of COVID-19. 
Like many authors, we adopt the classic epidemic SIR model, where the infection rate is the control variable. Its differential flatness property yields ele mentary closed-form formulae for open-loop social distancing scenarios, where, for instance, the increase of the number of uninfected people may be taken into account. Those formulae might therefore be useful to decision makers. A feedback loop stemming from model-free control leads to a remarkable robustness with respect to severe uncertainties and mismatches. Although an identification procedure is presented, a good knowledge of the recovery rate is not necessary for our control strategy.}


%This electronic document is a ÒliveÓ template. The various components of your paper [title, text, heads, etc.] are already defined on the style sheet, as illustrated by the portions given in this document.

\end{abstract}

%%%%%%%%%%%%%%%%%%%%%%%%%%%%%%%%%%%%%%%%%%%%%%%%%%%%%%%%%%%%%%%%%%%%%%%%%%%%%%%%
\newpage

\section{Introduction}
In two years an abundant mathematically oriented literature has been devoted to the worldwide COVID-19 pandemic. Some of the corresponding calculations had even a significant political impact (see, \textit{e.g.}, \cite{adam,quintana}). Note that in the field of mathematical epidemiology of infectious diseases the role of modeling human behavior became increasingly important in the last $15$ years. It gave birth to a novel research field named \textit{behavioral epidemiology} of infectious diseases: see, \textit{e.g.}, \cite{be2013,spva}. 

A novel control technique for improving the social distancing is presented here. This fundamental topic has already been tackled by many authors: see, \textit{e.g.},
\cite{al,ames,angulo,berger,bisiacco,bliman1,bliman2,bonnans,borri,charpentier,dilauro1,dilauro2,dias,efimov,gevertz,godera,greene,ianni,jing,kohler,mc,morato1,morato2,morgan,morris,osullivan,peni,pillonetto,sadeghi,sontag,stella,tsay}. Most of those papers are based on the famous \emph{SIR} (\emph{Susceptible-Infected-Recovered/Removed}) model, which goes back to 1927 (\cite{kermack}), or on some modifications of its \emph{compartments}. This communication is also using the SIR model:
\begin{itemize}
\item When, like in several papers, the \emph{infection rate} is the control variable, the SIR model is \emph{(differentially) flat} (\cite{ijc}). Remember that flatness-based control is one of the most popular model-based control setting, especially with respect to concrete applications: see, \textit{e.g.}, \cite{beltran,bonnabel,diwold,kogler,li,lorenz,miunske,richter,sahoo,sanchez,schorghuber,steckler,sekiguchi,tal,thounthong,tognon,zauner} for some recent publications. Note that flatness has already been utilized by \cite{hametner} for studying COVID-19 but with other purposes.
\item There are severe uncertainties: model mismatch, poorly known initial conditions, \dots We therefore close the loop around the reference trajectory via \emph{model-free} control, or \emph{MFC}, in the sense of \cite{mfc13,nicu}.
MFC, which is easy to implement, has already been illustrated in a number of practical situations. Some new contributions are listed here: \cite{gu,ismail,jin,kuruganti,lv,manzoni,mao,michel,mousavi,neves,sancak,sehili,srour,sun,xu0,xu1,wangc,wanga,wangb,zhang0,zhang1,zhou}. Let us single out here the excellent work by \cite{truong} on ventilators, which are related to COVID-19.
%\item  According to \cite{easy}, the \emph{recovery rate} in the SIR model is a \emph{rationally identifiable} parameter.
\end{itemize}
In order to be more specific consider a flat system with a single input $u$ and a single output $y$. Assume that $y$ is a flat output. Our strategy (see also \cite{villagra,cancer}) may be summarized as follows:
\begin{enumerate}
\item To any output reference trajectory $y^\star$ corresponds at once thanks to flatness an open-loop control $u^\star$.
\item Let $z$ be some measured output. Write $z^\star$ the corresponding reference trajectory. Set $u = u^\star + \Delta u$, where $\Delta u$ is the control of an \emph{ultra-local} local model (\cite{mfc13}). Its output $\Delta z = z - z^\star$ is the tracking error. Closing the loop via an \emph{intelligent controller} (\cite{mfc13}) permits to ensure local stability around $z^\star$ in spite of severe mismatches and disturbances.
\end{enumerate} 

%The above viewpoint was already successfully applied by \cite{21} for some computer experiments in oncology. It is further employed here in \emph{anti-angiogenesis}, \textit{i.e.}, for cancer treatments which were proposed at first by \cite{folkman}.  %Anti-angiogenesis aims to inhibit the spread of the vascular network necessary to support tumor growth during the vascular phase, so providing a way to control the heterogeneous and growth-unconstrained tumor population. A modeling via %nonlinear ordinary differential equations due to \cite{hahn} (see, \textit{e.g.}, \cite{ono09,onocell} for some modifications) has attracted a lot of attention, including in biomedical control: see, \textit{e.g.}, %%\cite{cacace1,cacace2,drexler,ergun,kovacs14,onocell}; \cite{scha} and references therein; \cite{sapi,szeles}. 

%The main features of our approach may be summarized as follows:
%\begin{itemize}
%\item Numerical simulations are derived easily with a low computing cost.
%\item An excellent robustness with respect to a poor knowledge of the parameters is clearly deduced.
%\end{itemize}
%\begin{remark}
%The above results confirm a previous paper on drug injections scheduling (\cite{21})  in a chemo- and immunotherapy via a cancer modeling due to \cite{ono12}, where a remarkable robustness with respect to uncertainty in drug delivery 
%(\cite{sharifi}) is obtained.
%\end{remark}

Our paper is organized as follows: 
\begin{itemize}
\item Section \ref{sir} shows that the SIR model, where the infection rate is the control variable, is flat and the population of recovered/removed individuals is a flat output; the recovery rate is identifiable in the sense of \cite{easy}.
\item Section \ref{flat} is devoted to a flatness-based control strategy, \textit{i.e.}, to a feedforward approach. Elementary closed-form of the control and state variables are easily derived. Various scenarios, where for instance the number of uninfected persons is increased, may thus be easily suggested to decision makers. 
\item Closing the loop via an intelligent proportional regulator, stemming from model-free control, is the subject of Section \ref{csm}. Computer simulations confirm an excellent robustness with respect to severe uncertainties.
\item A time-varying recovery rate is estimated in Section \ref{gamma} via \emph{algebraic estimation} methods (\cite{easy}). Techniques from Section \ref{csm} show however good performances if this rate is wrongly assumed to be constant.
\item Some suggestions for future investigations and someconcluding remarks may be found in Section \ref{conclusion}.
\end{itemize}

%%%%%%%%%%%%%%%%%%%%%%%%%
\section{Modeling issues}\label{sir}
\subsection{The SIR model}
The SIR model (see, \textit{e.g.}, \cite{weiss} for a nice introduction) reads:
\begin{equation}\label{SIR}
\begin{cases}
\dot{S} = -\beta IS \\
\dot{I} = \beta IS  - \gamma I \\
\dot{R} = \gamma I
\end{cases}
\end{equation}
$S$, $I$ and $R$, which are non-negative quantities, correspond respectively to the fractions of susceptible, infected and recovered/removed individuals in the population. We may set therefore
\begin{equation}\label{1}
S + I + R = 1
\end{equation}
$\beta$, $0 < \underline{\beta} \leq \beta \leq \overline{\beta}$, which is here the control variable,\footnote{Softening social distancing implies increasing $\beta (t)$.} and the parameter $\gamma > 0$ are respectively the infection and recovery rates.  
\subsection{Flatness}
Equations \eqref{SIR}-\eqref{1} show that System \eqref{SIR} is flat and that $R$ is a flat output (\cite{ijc}). The other system variables may be expressed as \emph{differential rational functions} of $R$, \textit{i.e.}, as rational functions of $R$ and its derivatives up to some finite order:
\begin{align}
\label{a}
I &= \frac{\dot{R}}{\gamma} \\ 
\label{b}
S &= 1 - R - \frac{\dot{R}}{\gamma} \\ 
\label{c}
\beta &= - \frac{\dot{S}}{IS} = \frac{1}{S}\left(\frac{\dot{I}}{I} + \gamma \right)
\end{align}
\begin{remark}
If $\gamma$ is not constant, but a differentiable function of time, Equations \eqref{a}-\eqref{b}-\eqref{c} remain valid: System \eqref{SIR} is still flat and $R$ is still a flat output. Equation \eqref{c} shows however that 
$\dot{\gamma}$ is needed.
\end{remark}

\subsection{An addendum on the SEIR model}
The \emph{SEIR} model (see, \textit{e.g.}, \cite{brauer}) is a rather popular extension of the SIR model:
\begin{equation}\label{SEIR}
\begin{cases}
\dot{S} = -\beta IS \\
\dot{E} = \beta IS  - \alpha E \\
\dot{I} = \alpha E - \gamma I \\
\dot{R} = \gamma I
\end{cases}
\end{equation}
where $\alpha > 0$ is an additional parameter. Equation \eqref{1} becomes
%in the above model we observe a fourth state variable $E$ that denotes the fraction of \textit{exposed subjects}, i.e. of subject for whom the disease is latent, and the additional parameter $\alpha$ that is the inverse of the average time of disease %latency. In other words $1/\alpha$ is the average time needed for the vital population installed in the host at the contagion  to become sufficiently large in order that the subject can infect Susceptible subjects.  
\begin{equation}\label{inv}
S + E + I + R = 1.
\end{equation}
Equations \eqref{SEIR}-\eqref{inv} show that the SEIR model is also flat and that $R$ is a flat output: %The other system variables may be expressed as \emph{differential rational functions} of $R$, \textit{i.e.}, as rational functions of $R$ and its derivatives up to some finite order:
\begin{equation*}
\left\lbrace
\begin{aligned}
\label{Iflat}
I &= \frac{\dot{R}}{\gamma} \\ 
%\label{Eflat}
E &= \frac{\dot{I}+\gamma I}{\alpha} = \frac{\ddot{R}+\gamma \dot{R}}{\gamma\alpha} \\ 
%\label{Sflat}
S &= 1 - R - I - E=  1 - R - \frac{\dot{R}}{\gamma} - \frac{\ddot{R}+\gamma \dot{R}}{\gamma\alpha}  \\ 
%\label{betaflat}
\beta &= - \frac{\dot{S}}{IS} 
\end{aligned}
\right.
\end{equation*}

\subsection{Identifiability of the recovery rate}
Equation \eqref{c} yields 
$$\gamma =  \beta S - \frac{\dot{I}}{I}$$ 
$\gamma$ is a differential rational function .of $R$ and $\beta$: It is thus \emph{rationally identifiable} (\cite{easy}). 
\begin{remark}
The above equation does not work for an identifiability purpose if $\gamma$ is time-varying: $\dot{\gamma}$ is sitting on its right hand-side.  If we assume that $I$ and $S$ are measured, Equation \eqref{b} yields
\begin{equation}\label{bb}
\gamma = \frac{\dot{I} - \beta IS}{I}
\end{equation}
$\gamma$ is still rationally identifiable with respect to $I$, $S$, $\beta$. It will be useful in Section \ref{gamma}. 
\end{remark}

\section{Flatness-based control}\label{flat}
\subsection{Preparatory calculations}
Set $$\boxed{I_{\rm reference}(t) = I_0 e^{-\lambda t}}$$ 
where $t \geq 0$, $0 \leq I_0 \leq 1$, and $\lambda \geq 0$ is some constant parameter. 
If we set $R(0) = 0$,  it yields 
{$$R_{\rm reference}(t) = \frac{\gamma I_0}{\lambda}(1 - e^{-\lambda t})$$
$$S_{\rm reference} (t) = 1-\frac{\gamma I_0}{\lambda}\left(1-e^{-\lambda t}\right)-I_0e^{-\lambda t}$$
and the open-loop control
$$\boxed{\beta_{\rm flat} (t)  = \frac{\gamma-\lambda}{1-\frac{\gamma I_0}{\lambda}\left(1-e^{-\lambda t}\right)-I_0e^{-\lambda t}}}$$
Thus 
\begin{equation}\label{rat}
\lim\limits_{t\to +\infty} \beta_{\rm flat} (t) = \frac{\lambda(\gamma-\lambda)}{\lambda-\gamma I_0}
\end{equation}
The following inequalities are staightforward:
\begin{equation}\label{ineq1}
 \gamma I_0 < \lambda < \gamma
\end{equation}
$\lambda < \gamma$ follows from $\beta > 0$; $\gamma I_0 <  \lambda$ follows from 
\begin{equation}\label{susc}
\lim\limits_{t\to +\infty}S(t) = 1-\frac{\gamma I_0}{\lambda} = S(\infty) >0 
\end{equation}
Introduce the more or less precise quantity $\beta_{\rm accept}$, where $\underline{\beta} < \beta_{\rm accept} < \overline{\beta}$. It stands for the ``harshest'' social distancing protocols which are ``acceptable'' in the long run. Equation \eqref{rat} yields therefore 
$$\frac{\lambda(\gamma-\lambda)}{\lambda-\gamma I_0} = \beta_{\rm accept}$$
The positive root of the corresponding quadratic algebraic equation 
$\lambda^2 + (\beta_{\rm accept} - \gamma)\lambda - \gamma I_0 \beta_{\rm accept} = 0$ 
is
$$
\boxed{\lambda_{\rm accept} = \frac{{\gamma-\beta_{\rm accept}+\sqrt{\Delta_{\rm accept}}}}{2}}
$$
where $\Delta_{\rm accept} = (\gamma-\beta_{\rm accept})^2+4\gamma I_0\beta_{\rm accept} \geq 0$.
The fundamental inequality
\begin{equation*}\label{lambda}
\gamma I_0 < \lambda_{\rm accept} < \gamma
\end{equation*}
follows from 
$$\lim\limits_{\lambda \downarrow \gamma I_0} \frac{\lambda(\gamma-\lambda)}{\lambda-\gamma I_0} = + \infty, \quad \lim\limits_{\lambda \uparrow \gamma} \frac{\lambda(\gamma-\lambda)}{\lambda-\gamma I_0} = 0$$
Equation \eqref{susc} leads to the notation 
$$S_{\rm accept} (\infty) = 1-\frac{\gamma I_0}{\lambda_{\rm accept} }$$ 
The inequality 
$$S (\infty) < S_{\rm accept} (\infty) \quad {\rm if} \quad  \lambda < \lambda_{\rm accept} $$
demonstrates that the proportion of uninfected people decreases if the social distancing obligations are relaxed.
\subsection{Two computer experiments}\label{exp}
Set $\gamma = 0.1$, $\beta_{\rm accept} = 0.22$. Figure \ref{ol} displays the open-loop evolutions of $\beta$, $I$, $S$ when $\lambda = \lambda_{\rm accept}$. Those behaviors are quite satisfactory.

\section{Model-free control}\label{csm}
\subsection{Ultra-local model}
Set $\Delta I (t) = I(t) - I_{\rm reference} (t)$, $\beta (t) = \beta_{\rm flat} (t) + \Delta \beta (t)$. In order to take into account the various uncertainties, introduce the \emph{ultra-local} model (\cite{mfc13})
\begin{equation}\label{ul}
\frac{d}{dt} \Delta I = F + \frak{a} \Delta \beta
\end{equation}
\begin{itemize}
\item The function $F$, which is data-driven, subsumes the poorly known structures and disturbances. 
\item The parameter $\frak{a}$, which does not need to be precisely determined, is chosen such that the three terms in Equation \eqref{ul} are of the same magnitude.
\item {\small $F_{\rm est} = - \frac{6}{\tau^{3}} \int_{t-\tau}^t \left( (t-2\sigma)\Delta I(\sigma) + \frak{a} \sigma(\tau - \sigma)\Delta \beta(\sigma)\right)d\sigma$}, where $\tau > 0$ is ``small'', gives a real-time estimate, which in practice is implemented via a digital filter.
\end{itemize}
\subsection{Intelligent proportional controller}
Introduce (\cite{mfc13}) the \emph{intelligent proportional controller}, or \emph{iP},
\begin{equation}\label{ip}
\boxed{\Delta \beta = - \frac{F_{\rm est} + K_P \Delta I}{\frak{a}}}
\end{equation}
where $K_P$ is a tuning gain. Equations \eqref{ul} and \eqref{ip} yield $$\frac{d}{dt} \Delta I + K_P \Delta I = F - F_{\rm est}$$ 
Set $K_P > 0$. Then $\lim\limits_{t\to +\infty} \Delta I(t) \approx  0$ if the estimate $F_{\rm est}$ is ``good,'' \textit{i.e.}, if $F - F_{\rm est}$ is ``small.'' Local stability is ensured.
\begin{remark}
When compared to classic PIs and PIDs (see, \textit{e.g.}, \cite{astrom}), the gain tuning of the iP is straightforward.
\end{remark}
\subsection{Computer experiments}
The sampling time interval is $2$ hours. In Equations \eqref{ul} and \eqref{ip}, $\frak{a} = 0.1$, $K_P = 1$. Figure \ref{sc1} displays excellent results in spite of errors on initial conditions and of
the fuzzy character of any measurement of the social distancing. This fuzziness  is expressed here by an additive corrupting white Gaussian noise $\mathcal{N}(0,5.10^{-3})$ on $\beta$.  

%Dans un sixième scénario, on considère d'une part la condition initiale mal connue (voir Fig. \ref{BF6}-(b) $I_0=0.1\neq0.13$) et d'autre part que les mesures de distanciation ne sont pas appliquées précisément. Ceci est simulé au moyen d'un %bruit, $\mathcal{N}(0,5.10^{-3})$, ajouté à la commande $\beta$. Comme le montre la figure \ref{BF6}, les résultats restent excellents.
\section{On the recovery rate $\gamma$}\label{gamma}
Assume now that $\gamma$ is a differentiable time function. Equation \eqref{bb} yields the algebraic estimator
\begin{equation}\label{estim}
\gamma_{\rm est} = \frac{[\dot I]_{\rm est} - \beta I S}{I}
\end{equation}
where $[\dot I]_{\rm est}$ is an estimate of $\dot{I}$ obtained along the lines developed by \cite{mboup} and \cite{othmane} for {\em algebraic differentiators}. Figure \ref{BF4}-c displays excellent results. The flatness-based computer experiments is achieved as in Section \ref{exp}, \textit{i.e.}, $\gamma = 0.1$ is assumed to be constant. Lack of space prevents us from examining more realistic situations. Closing the loop via model-free control yields as demonstrated in Figures \ref{BF4}-a-b a satisfactory behavior. Is the exact knowledge of the recovery rate unimportant?

\section{Conclusion}\label{conclusion}  

\cite{casella} questions the  relevance and usefulness of such control-theoretic considerations for non-pharmaceu\-tical mitigation policies against COVID-19. We certainly do not claim  to set aside those objections in this preliminary short study. The  combination however of flatness-based and model-free controls presents nevertheless some major advantages as demonstrated here and by \cite{villagra} and \cite{cancer}. 
 
An extra theoretical effort must be made in order  to  design control strategy as close as possible to the real epidemic control enacted by Public Health authorities. Summarizing, we consider this results proposed in this work as a theoretical ideal framework, to be filled with a more realistic picture: an implementable non-pharmaceutical control strategy. Preliminary results, which we recently obtained, indicate that the methodology here proposed is in the right direction (see \cite{join}).

%\newpage                           

%{\scriptsize                                                   

\begin{figure*}[!h]
\centering
\subfigure[$\beta$]{\rotatebox{-0}{\epsfig{figure=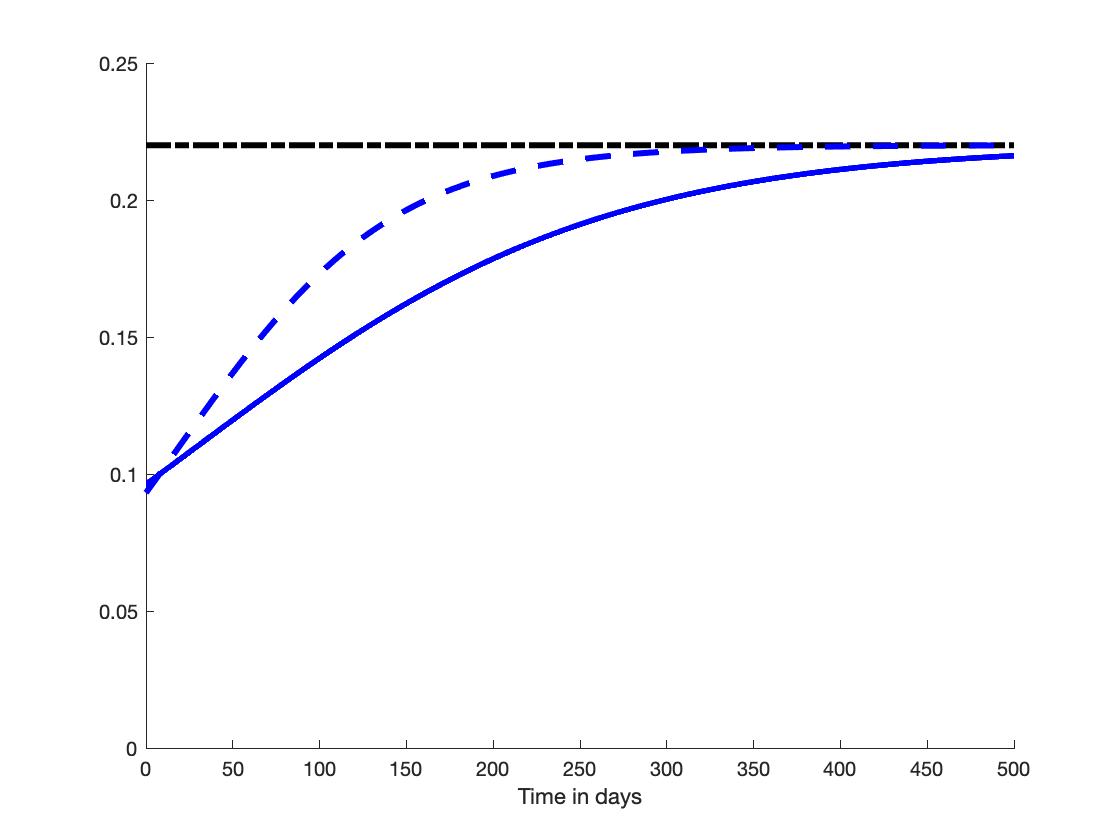,width=0.3\textwidth}}}
\subfigure[$I$]{\rotatebox{-0}{\epsfig{figure=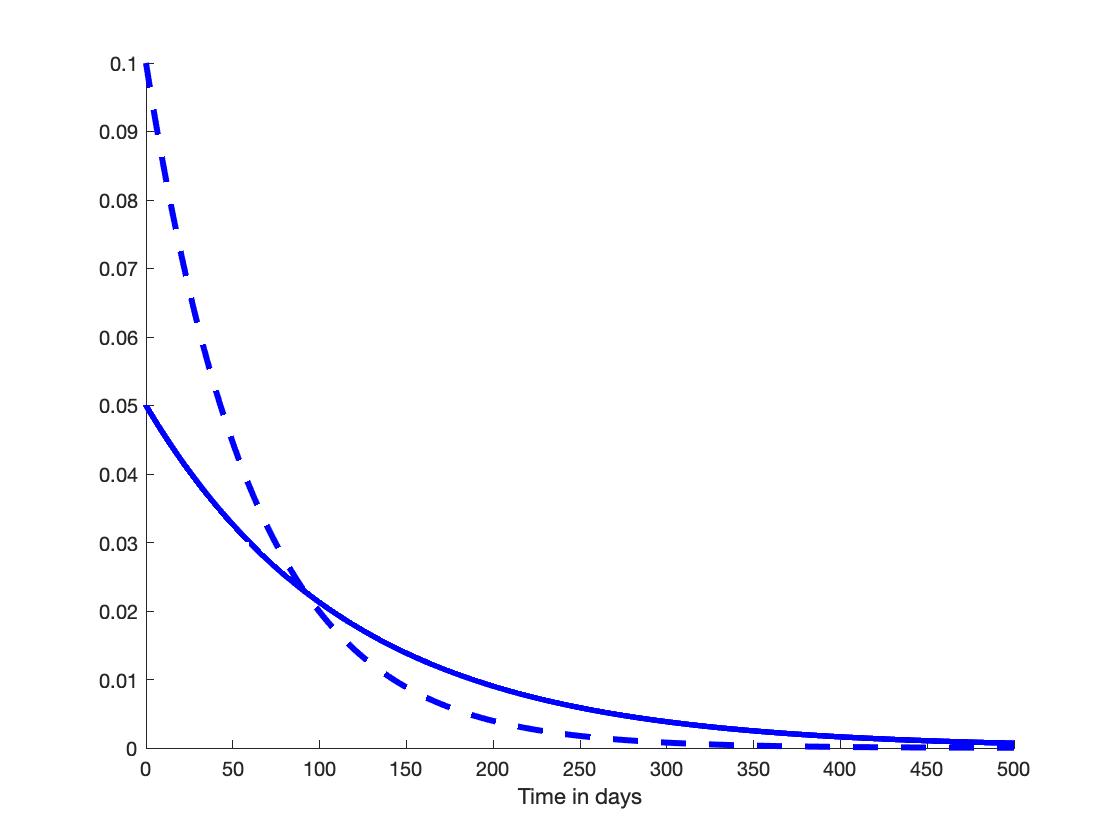,width=0.3\textwidth}}}
\subfigure[$S$]{\rotatebox{-0}{\epsfig{figure=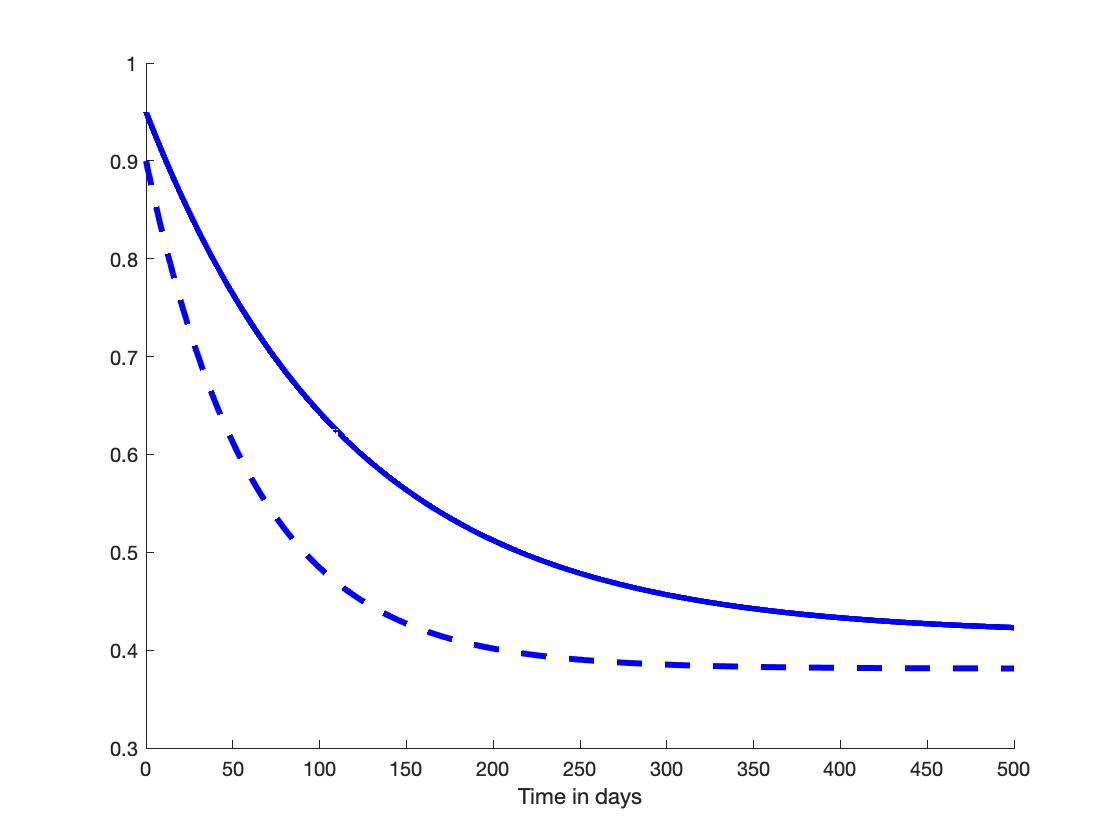,width=0.3\textwidth}}}
\caption{Open loop control strategy. Trajectories corresponding to two distinct initial conditions for the infectius  $I_0 = 0.05$ (single-dashed curves: -) and $I_0 =  0.1$ (double-dashed curves: - -). Left panel: plot of the transition rate $\beta(t)$; central panel: plot of the infectious fraction $I(t)$; right panel: plot of the fraction of susceptible subjects $S(t)$.}\label{ol}
\end{figure*}

\begin{figure*}[!h]
\centering
\subfigure[$\beta$]{\rotatebox{-0}{\epsfig{figure=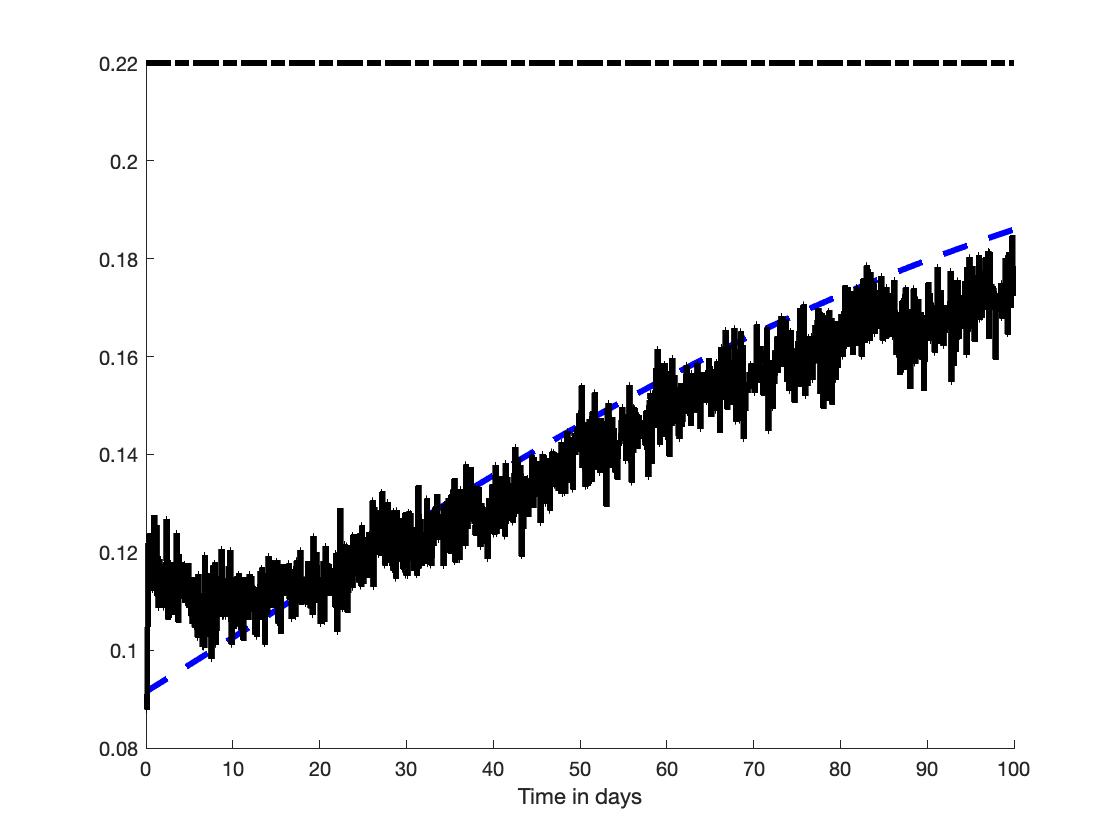,width=0.3\textwidth}}}
\subfigure[$I$]{\rotatebox{-0}{\epsfig{figure=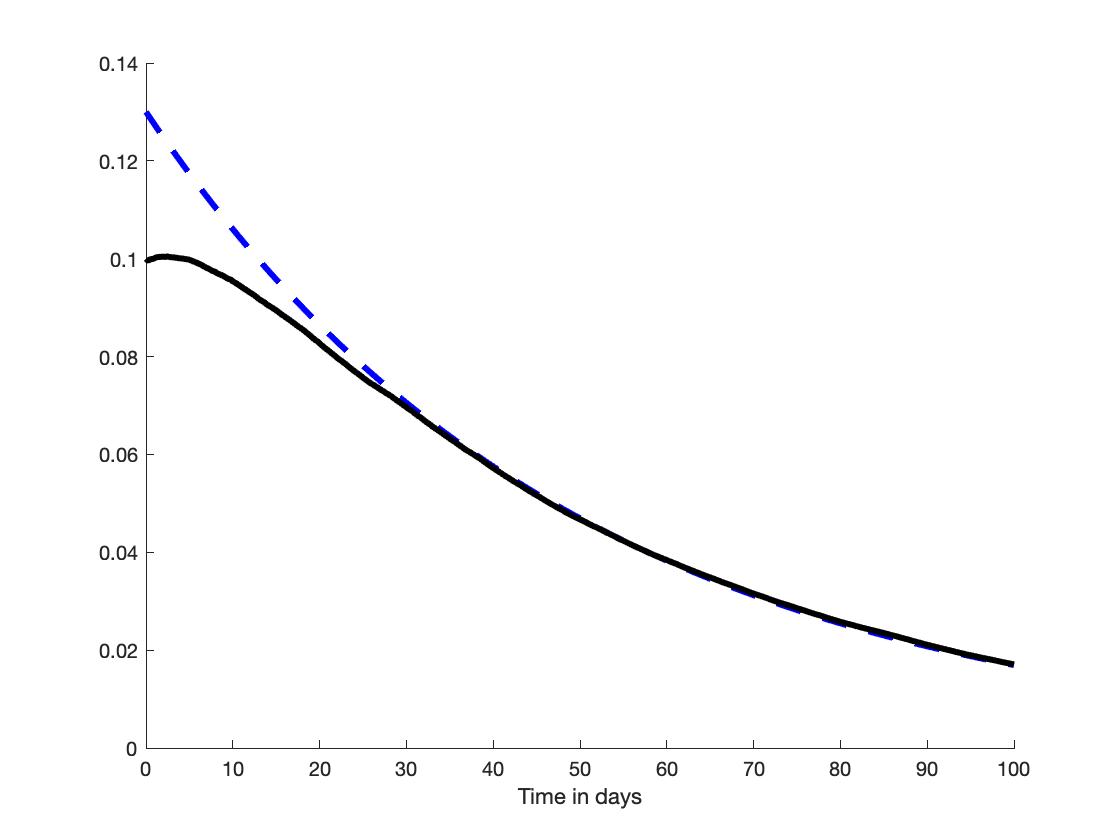,width=0.3\textwidth}}}
\subfigure[$S$]{\rotatebox{-0}{\epsfig{figure=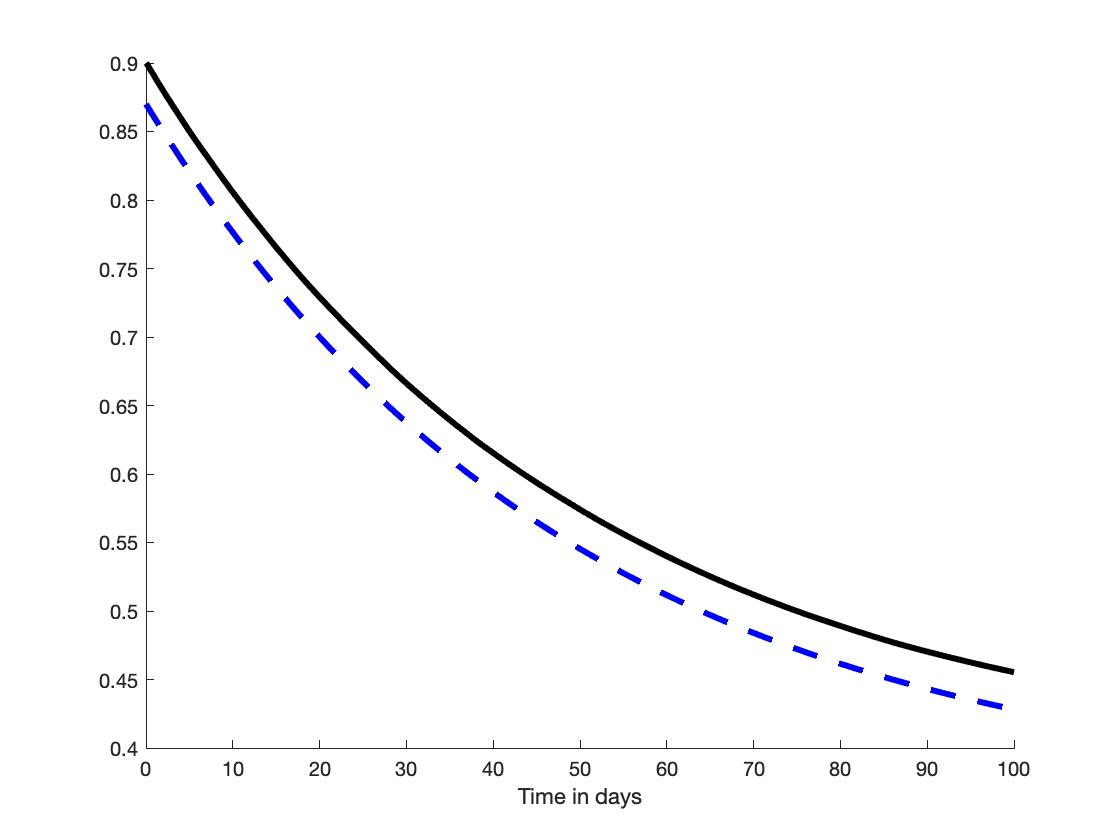,width=0.3\textwidth}}}
%\subfigure[$\gamma$ (- .) et ses estimations $\gamma_{e1}$ (bleu --) $\gamma_{e2}$ (rouge - -)]{\rotatebox{-0}{\epsfig{figure=F5G,width=0.5\textwidth}}}
\caption{Effect of both errors on initial conditions and of the fuzziness of measurements of social distancing. In all panels, dashed blue line represent the reference trajectories. Left panel: plot of the transition rate $\beta(t)$; central panel: plot of the infectious fraction $I(t)$; right panel: plot of the fraction of susceptible subjects $S(t)$.}\label{sc1}%Error on initial conditions and fuzzy $\beta$ -- blue(- -): reference trajectory t)$?.
%6)      Figure 3: it is better to put the figure on gamma as panel (a), as it is also first cited in the text. I suggest to use a version with two upper panels and two lower panels. Or to split figure 3 I two figures: one for $\gamma$ and the other for the %MFC.
%7)      In the list of works adopting }\label{sc1}
\end{figure*}

\begin{figure*}[!h]
\centering
\subfigure[{$\beta$ -- (blue - -): reference trajectory}]{\rotatebox{-0}{\epsfig{figure=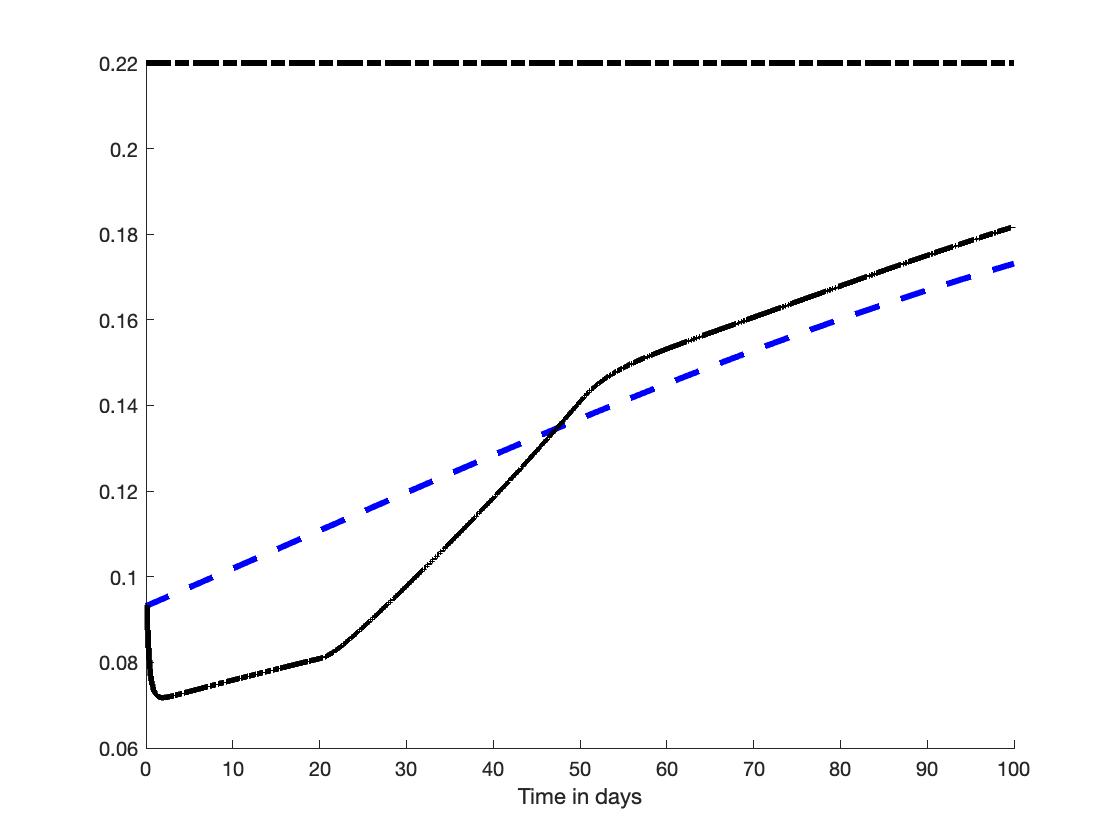,width=0.3\textwidth}}}
\subfigure[{$I$ -- (blue - -): reference trajectory}]{\rotatebox{-0}{\epsfig{figure=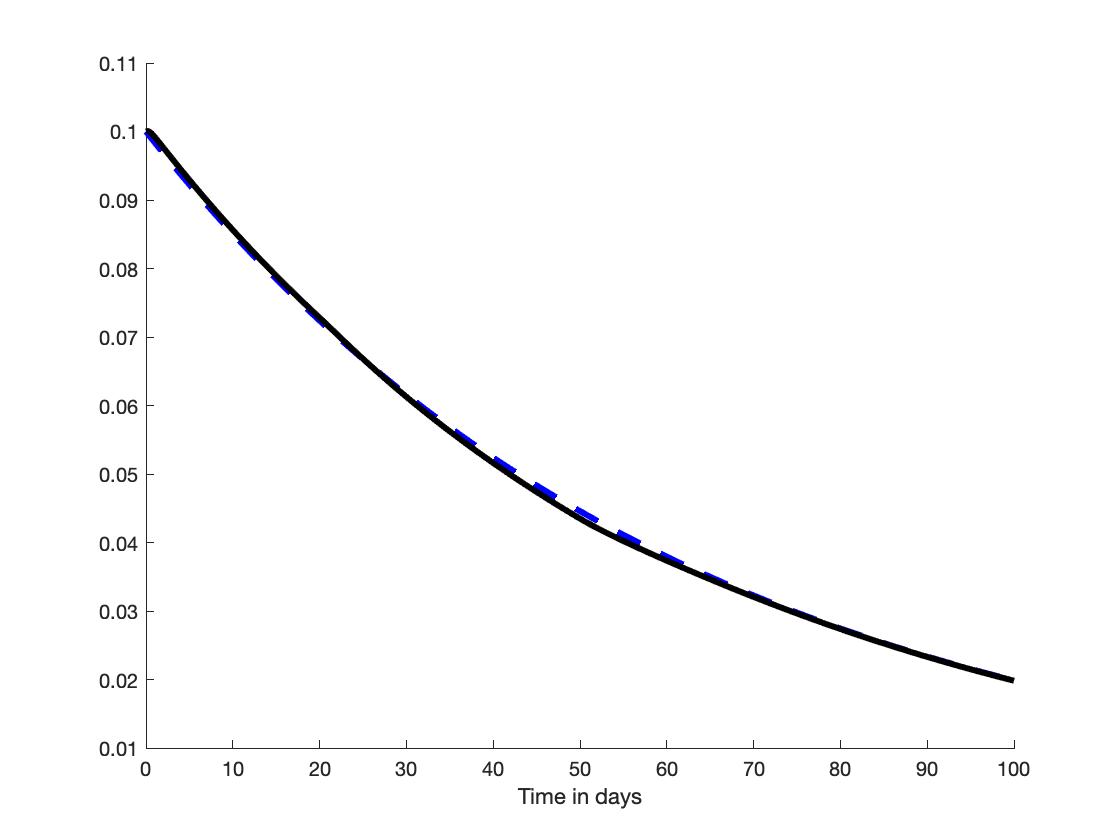,width=0.3\textwidth}}}
\subfigure[{$S$ -- (blue - -): reference trajectory}]{\rotatebox{-0}{\epsfig{figure=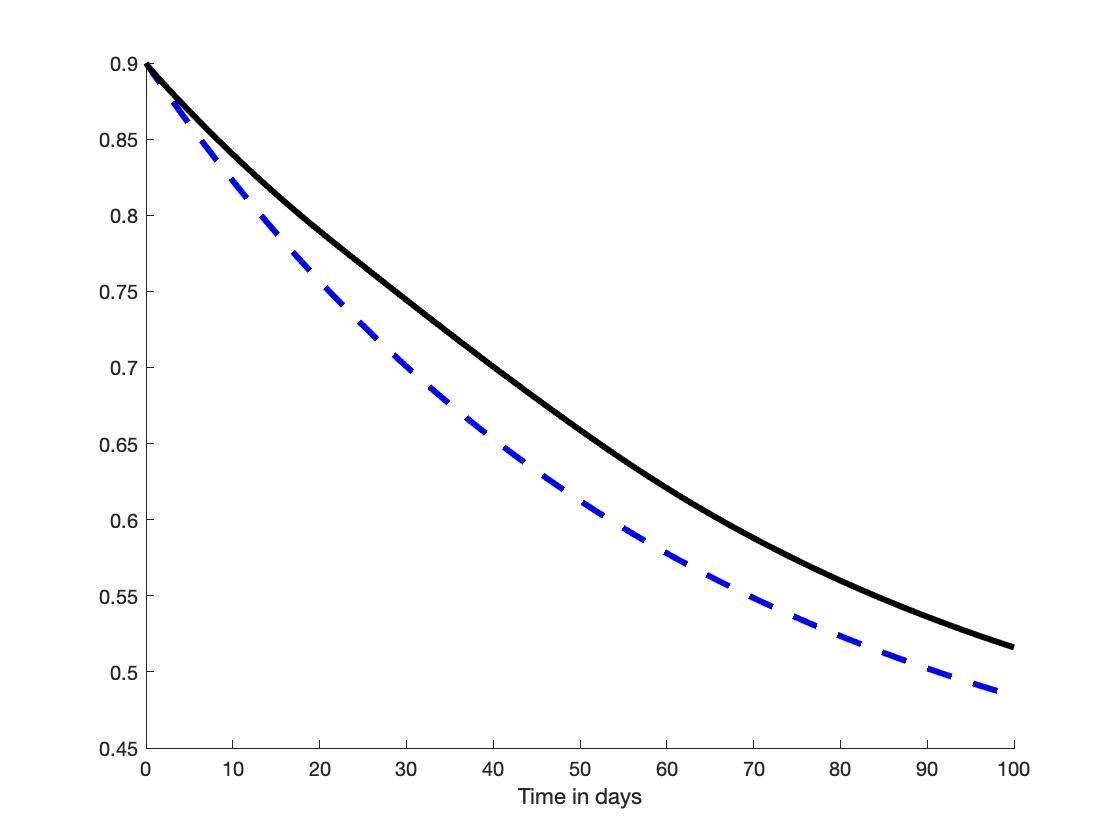,width=0.3\textwidth}}}
%Mathematical Modeling and Optimal Control of Complex Epidemiological Networks
\subfigure[{\scriptsize $\gamma$ (- -) and $\gamma_{\rm est}$ (blue --)}]{\rotatebox{-0}{\epsfig{figure=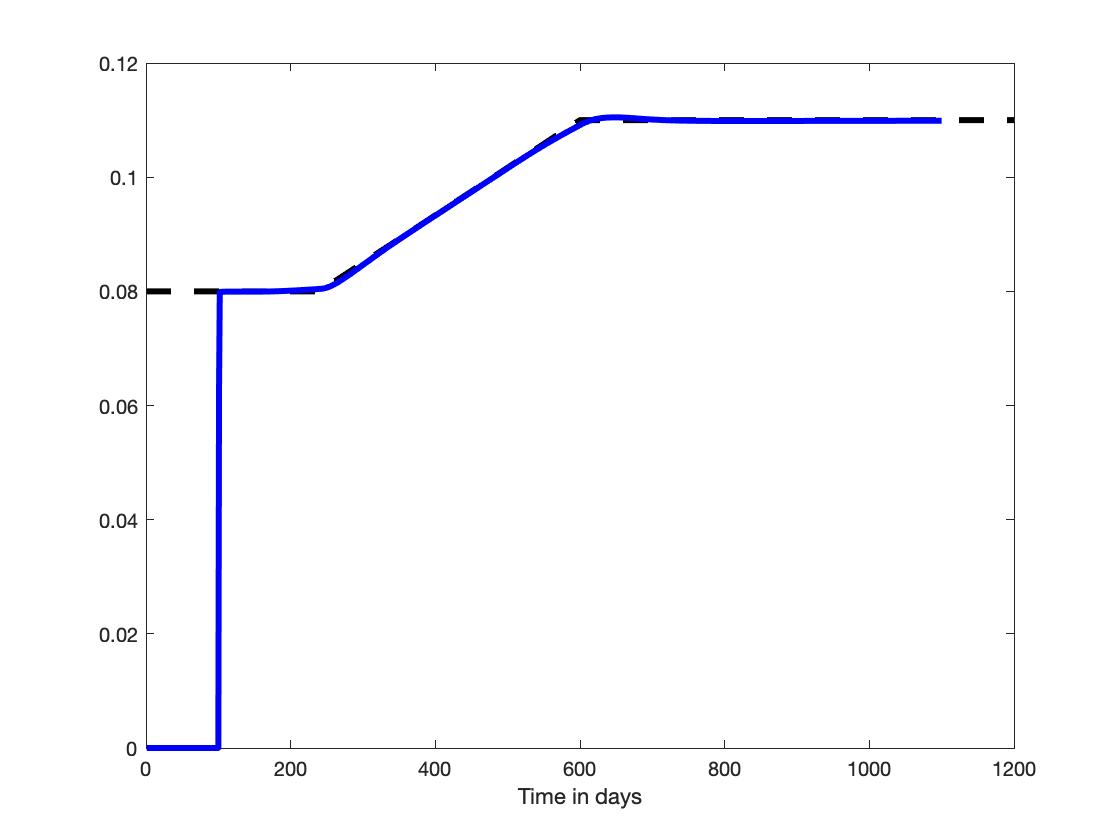,width=0.25\textwidth}}}
\caption{Impact of the estimation of the time-varying recovery rate $\gamma$.}\label{BF4}
\end{figure*}


\begin{thebibliography}{xx}  % you can also add the bibliography by hand
{%\scriptsize
\bibitem{adam}
Adam D. (2020). 
\newblock Special report: the simulations driving the world's response to COVID-19.
\emph{Nature}, 580, 316-318.

\bibitem{al}
Al-Radhawi M.A., Sadeghi M., Sontag E.D. (2022).
\newblock Long-term regulation of prolonged epidemic outbreaks in large populations via adaptive control: A singular perturbation approach.
\newblock \emph{IEEE Contr. Syst. Lett.}, 6, 578-583.

\bibitem{ames}
Ames A.Z., Moln\'{a}r T.G., Singletary A.W., Orosz G. (2020).
\newblock Safety-critical control of active interventions for COVID-19 mitigation.
\newblock \emph{IEEE Access}, 8, 188454-188474.

\bibitem{angulo}
Angulo M.T., Casta$\tilde{\rm n}$os F., Moreno-Morton R., Velasco-Hern\'{a}ndez J.X., Moreno J.A. (2021).
\newblock A simple criterion to design optimal non-pharmaceutical interventions for mitigating epidemic outbreaks.
\newblock \emph{J. Roy. Soc. Interface}, 18, 20200803.

\bibitem{astrom}
{\AA}str\"om K.J., Murray R.M. (2008).
\newblock \emph{Feedback Systems: An Introduction for Scientists and Engineers}. 
\newblock Princeton University Press.


\bibitem{beltran}
Beltran-Carbajal F., Tapia-Olvera R., Valderrabano-Gonzalez A., Yanez-Badillo H., Rosas-Caro J.C., Mayo-Maldonado J.C. (2021).
\newblock Closed-loop online harmonic vibration estimation in DC electric motor systems.
\newblock \emph{Appl. Math. Model.}, 94, 460-481.

\bibitem{berger}
Berger T. (2022).
\newblock Feedback control of the COVID-19 pandemic with guaranteed non-exceeding ICU capacity.
\newblock \emph{Syst. Contr. Lett.}, 160, 105111.

\bibitem{bisiacco}
Bisiacco M., Pillonetto G. (2021).
\newblock COVID-19 epidemic control using short-term lockdowns for collective gain.
\newblock \emph{Ann. Rev. Contr.}, 52, 573-586.

\bibitem{bliman1}
Bliman P.-A., Duprez M. (2021). 
\newblock How best can finite-time social distancing reduce epidemic final size?
\newblock \emph{J. Theoret. Biol.}, 511, 110557.

\bibitem{bliman2}
Bliman P.-A., Duprez M., Privat Y., Vauchelet N. (2021).
\newblock Optimal immunity control and final size minimization by social distancing for the SIR epidemic model.
\newblock \emph{J. Optim. Theory App.}, 189, 408-436.

\bibitem{bonnabel}
Bonnabel S., Clayes X. (2020).
\newblock The industrial control of tower cranes: An operator-in-the-loop approach.
\newblock \emph{IEEE Contr. Syst. Magaz.}, 40,  27-39.

\bibitem{bonnans}
Bonnans J.F., Gianatti J. (2020).
\newblock Optimal control techniques based on infection age for the study of the COVID-19 epidemic.
\newblock  \emph{Math. Model. Nat. Phenom.}, 15, 48.

\bibitem{borri}
Borri A., Palumbo P., Papa F., Possieri C. (2021).
\newblock Optimal design of lock-down and reopening policies for early-stage epidemics through SIR-D models
\newblock \emph{Annu. Rev. Contr.} 51, 511-524.

\bibitem{brauer}
Brauer F., Castillo-Chavez C. (2012). 
\newblock\emph{Mathematical Models in Population Biology and Epidemiology} (2nd ed.).
\newblock Springer.

\bibitem{casella}
Casella F. (2021).
\newblock Can the COVID-19 epidemic be controlled on the basis of daily test reports?
\newblock \emph{IEEE Contr. Syst. Lett.}, 5, 1079-1084.

\bibitem{charpentier}
Charpentier A., Elie R., Lauri\`{e}re`M., Tran V.C. (2020).
\newblock COVID-19 pandemic control: balancing detection policy and lockdown intervention ICU sustainability.
\newblock \emph{Math. Model. Nat. Phenom.}, 15, 57.

\bibitem{dias}
Dias S., Queiroz K., Araujo A. (2022).
\newblock Controlling epidemic diseases based only on social distancing level.
\newblock \emph{J. Contr. Autom. Electr. Syst.}, 33, 8-22.

\bibitem{dilauro1}
Di Lauro F., Kiss I.Z., Della Santina C. (2021a).
\newblock Optimal timing of one-shot interventions for epidemic control.
\newblock \emph{PLoS Comput. Biol.}, 17, e1008763.

\bibitem{dilauro2}
Di Lauro F., Kiss I.Z., Della Santina C. (2021b).
\newblock Covid-19 and flattening the curve: A feedback control perspective.
\newblock   \emph{IEEE Contr. Syst. Lett.}, 5, 1435-1440.

\bibitem{diwold}
Diwold J., Kolar B., Markus Sch\"{o}berl M. (2022).
\newblock Discrete-time flatness-based control of a gantry crane.
\newblock \emph{Contr. Engin. Pract.}, 119, 104980.


\bibitem{efimov}
Efimov D., Ushirobira R. (2021).
\newblock On an interval prediction of COVID-19 development based on a SEIR epidemic model.
\emph{Ann. Rev. Contr.}, 51, 477-487.


\bibitem{mfc13}
Fliess M., Join C. (2013).
\newblock Model-free control.
\newblock\emph{Int. J. Contr.}, 86, 2228-2252.

\bibitem{nicu}
Fliess M., Join C. (2021). 
\newblock An alternative to proportional-integral and proportional-integral-derivative regulators: Intelligent proportional-derivative regulators. 
\newblock \emph{Int. J. Robust Nonlin. Contr.} \newline {\tt  https://doi.org/10.1002/rnc.5657}

%\bibitem{arxiv}
%Fliess M., Join C. (2021b)
%\newblock Elementary formulae for social distancing scenarios: Application to COVID-19 mitigation via feedback control.
%\newblock {\tt arXiv:2110.01712}

\bibitem{cancer}
Fliess M., Join C., Moussa K., Djouadi S.M., Alsager M.W. (2021). 
\newblock Toward simple in silico experiments for drugs administration in some cancer treatments.
\newblock \emph{IFAC PapersOnLine}, 54-15, 245-250.

\bibitem{easy}
Fliess M., Join C., Sira-Ram\`{\i}rez H. (2008).
\newblock Non-linear estimation is easy.
\newblock \emph{Int. J. Model. Identif. Contr.}, 4, 12-27. 

\bibitem{ijc}
Fliess M., L\'{e}vine J., Martin P., Rouchon P. (1995).
\newblock Flatness and defect of non-linear systems: introductory theory and examples.
\newblock\emph{Int. J. Contr.}, 61, 1327-1361.

\bibitem{gevertz}
Gevertz J.L., Greene J.M., Sanchez-Tapia C.H., Sontag E.D. (2021).
\newblock A novel COVID-19 epidemiological model with explicit susceptible and asymptomatic isolation compartments reveals unexpected consequences of timing social distancing.
\newblock \emph{J. Theoret. Biol.}, 510, 110539.

\bibitem{godera}
Godara P., Herminghaus S., Heidemann K.M. (2021). 
\newblock A control theory approach to optimal pandemic mitigation. 
\newblock \emph{PLoS ONE}, 16, e0247445.

\bibitem{greene}
Greene J.M., Sontag E.D. (2021).
\newblock Minimizing the infected peak utilizing a single lockdown: a technical result regarding equal peak.
\newblock \emph{MedRxiv}. \newline {\tt https://doi.org/10.1101/2021.06.26.21259589}

\bibitem{gu}
Gu J., Li H., Zhang H., Pan C., Luan Z. (2021).
\newblock Cascaded model-free predictive control for single-phase boost power factor correction converters.
\newblock \emph{Int. J. Robust Nonlinear Contr.}, 31, 5016-5032.


\bibitem{hametner}
Hametner C., Kozek M., B\"{o}hler L., Wasserburger A., Peng Du Z., K\"{o}lbl R., Bergmann M., Bachleitner-Hofmann T., Jakubek S. (2021).
\newblock Estimation of exogenous drivers to predict COVID-19 pandemic using a method from nonlinear control theory.
\newblock \emph{Nonlin. Dyn.},  106, 1111-1125.


\bibitem{ianni}
Ianni A.,  Rossi N. (2021).
\newblock SIR-PID: A proportional-integral-derivative controller for COVID-19 outbreak containment.
\newblock \emph{Physics}, 3, 459-472.

\bibitem{ismail}
Ismail A., Noura H., Harmouch F., Harb Z. (2021).
\newblock Design and control of a neonatal incubator using model-free control.
\newblock \emph{29th Medit. Conf. Contr. Automat.}, Puglia.

\bibitem{jin}
Jin N., Chen M., Guo L., Li Y., Chen Y. (2021). 
\newblock Double-vector model-free predictive control method for voltage source inverter with visualization analysis.
\newblock  \emph{IEEE Trans. Indust. Electron.} {\tt doi: 10.1109/TIE.2021.3128905}

\bibitem{jing}
Jing M., Yew Ng K., Mac Namee B., Biglarbeigi P., Brisk R., Bond R., Finlay D., McLaughlin J. (2021).
\newblock COVID-19 modelling by time-varying transmission rate associated with mobility trend of driving via Apple Maps.
\newblock \emph{J. Biomed. Informat.}, 122, 103905.

\bibitem{join}
Join C., d'Onofrio A., Fliess M. (2022).
\newblock Toward realistic social distancing policies via advanced feedback control.
\newblock \emph{Submitted}.

\bibitem{kermack}
Kermack W.O., McKendrick A.G. (1927).
\newblock A contribution to the mathematical theory of epidemics.
\newblock \emph{Proc. Royal Soc. London Ser. A}, 115, 700-721.

\bibitem{kogler}
Kogler H., Ladner K., Ladner P. (2022). 
\newblock Flatness-based control of a closed-circuit hydraulic press.
\newblock In: Irschik H., Krommer M., Matveenko V.P., Belyaev A.K. (Eds) \emph{Dynamics and Control of Advanced Structures and Machines. Advanced Structured Materials}, pp. 111-121. 
\newblock Springer.

\bibitem{kohler}
K\"{o}hler J., Schwenkel L., Koch A., Berberich J., Pauli P., Allg\"{o}wer F. (2021).
\newblock Robust and optimal predictive control of the COVID-19 outbreak.
\newblock \emph{Ann. Rev. Contr.}, 51, 525-539.


\bibitem{kuruganti}
Kuruganti T., Olama M., Dong J., Xue Y., Winstead C., Nutaro J., Djouadi S., Bai L., Augenbroe G., Hill J. (2021).
\newblock \emph{Dynamic Building Load Control to Facilitate High Penetration of Solar Photovoltaic Generation}.
\newblock {Tech. Rep. ORNL/TM-2021/2112}, Oak Ridge National Lab.

\bibitem{li}
Li X., Wang Y., Guo X., Cui X., Zhang S., Li Y.(2021).
\newblock An improved model-free current predictive control method for SPMSM drives.
\newblock \emph{IEEE Access}, {\tt DOI: 10.1109/ACCESS.2021.3115782}

\bibitem{lorenz}
Lorenz-Meyer M.N.L., Menzel R., Dadzis K., Nikiforova A., Riemann H. (2020).
\newblock Lumped parameter model for silicon crystal growth from granulate crucible.
\newblock \emph{Cryst. Res. Techno.}, 55, 2000044.

\bibitem{lv}
Lv M., Gao S., Wei Y., Zhang D., Qi H., Wei Y.  (2022).
\newblock Model-free parallel predictive torque control based on ultra-local model of permanent magnet synchronous machine. 
\newblock \emph{Actuators}, 11, 31.

\bibitem{mao}
Mao J., Li H., Yang L., Zhang H., Liu L., Wang X., Tao J. (2021).
\newblock Non-cascaded model-free predictive speed control of SMPMSM drive system.
\newblock \emph{IEEE Trans. Energ. Convers.}, {\tt doi: 10.1109/TEC.2021.3090427}

\bibitem{be2013} 
Manfredi P., d'Onofrio A. (Eds) (2013). 
\newblock \emph{Modeling the Interplay Between Human Behavior and the Spread of Infectious Diseases}. 
\newblock Springer.

\bibitem{manzoni} 
 Manzoni E., Rampazzo M. (2021). 
 \newblock Automatic regulation of anesthesia via ultra-local model control via ultra-local model control. 
 \newblock \emph{IFAC PapersOnLine}, 54-15, 49-54. 

\bibitem{mboup}
Mboup M., Join C., Fliess M. (2009).
\newblock Numerical differentiation with annihilators in noisy environment.
\newblock \emph{Numer. Algor.}, 50, 439-467.

\bibitem{mc}
McQuade S.T., Weightman R., Merrill N.J., Yadav A., Tr\'{e}lat E.,  Allred S.R., Piccoli B. (2021).
\newblock Control of COVID-19 outbreak using an extended SEIR model.
\newblock \emph{Math. Model. Meth. Appl. Sci.}, 31, 2399-2424.

\bibitem{michel}
Michel L., Silva C.J., Torres D.F.M. (2022).
\newblock Model-free based control of a HIV/AIDS prevention model.
\newblock  \emph{Math. Biosci. Engin.}, 19, 759-774.

\bibitem{miunske}
Miunske T. (2020).
\newblock  \emph{Ein szenarienadaptiver Bewegungsalgorithmus für die L\"{a}ngsbewegung eines vollbeweglichen Fahrsimulators}.
\newblock Springer. 

\bibitem{morato1}
Morato M.M., Bastos S.B., Cajueiro D.O., Normey-Rico J.E. (2020a).
\newblock An optimal predictive control strategy for COVID-19 (SARS-CoV-2) social distancing policies in Brazil.
\newblock \emph{Annual Rev. Contr.}, 50, 417-431.

\bibitem{morato2}
Morato M.M., Pataro I.M.L., Americano da Costa M.V., Normey-Rico J.E. (2020b).
\newblock A parametrized nonlinear predictive control strategy for relaxing COVID-19 social distancing measures in Brazil.
\newblock \emph{ISA Trans.}, \newline {\tt https://doi.org/10.1016/j.isatra.2020.12.012}

\bibitem{morgan}
Morgan A.L.K., Woolhouse M.E.J., Medley G.F., van Bunnik B.A.D. (2021). 
\newblock Optimizing time-limited non-pharmaceutical interventions for COVID-19 outbreak control. 
\emph{Phil. Trans. Roy. Soc. B}, 376, 20200282. 

\bibitem{morris}
Morris D.H., Rossine F.W., Plotkin J.B., Levin S.A. (2021).
\newblock Optimal, near-optimal, and robust epidemic control.
\newblock \emph{Communic. Phys.}, 4, 78.

\bibitem{mousavi}
Mousavi  M.S., Davari  S.A., Nekoukar V., Garcia C., Rodriguez J. (2021).
\newblock Model-free finite set predictive voltage control of induction motor.
\newblock \emph{12th Power Electron. Drive Syst. Techno. Conf.}, Tabriz.

\bibitem{neves}
das Neves P.G., Augusto Ang\'{e}lico B.A. (2021).
\newblock  Model-free control of mechatronic systems based on algebraic estimation.
\newblock \emph{Asian J. Contr.}, {\tt https://doi.org/10.1002/asjc.2596}

\bibitem{osullivan}
O'Sullivan D., Gahegan M., Exeter  D.J., Adams B. (2020).
\newblock Spatially explicit models for exploring COVID-19 lockdown strategies.
\newblock \emph{Trans. GIS}, 24, 967-1000.

\bibitem{othmane}
Othmane A., Rudolph J., Mounier H. (2021).
\newblock Systematic comparison of numerical differentiators and an application to model-free control.
\newblock \emph{Europ. J. Contr.}, 62, 113-119.

\bibitem{pillonetto}
Pillonetto G., Bisiacco M., Pal\`{u} G., Cobelli C. (2021).
\newblock Tracking the time course of reproduction number and lockdown's effect on human behaviour during SARS-CoV-2 epidemic: nonparametric estimation.
\newblock \emph{Sci. Rep.}, 11, 9772.

\bibitem{peni}
P\'{e}ni T., Csutak B., Szederk\'{e}nyi G., R\"{o}st G. (2020).
\newblock Nonlinear model predictive control with logic constraints for COVID-19 management.
\newblock  \emph{Nonlin. Dyn.}, 102, 1965-1986.


\bibitem{quintana}
Quintana I.O., Rosenstock S., Klein C. (2021).
\newblock The coordination dilemma for epidemiological modelers.
\newblock \emph{Biol. Philo.}, 36, 54.

\bibitem{richter}
Richter H., Warner H. (2021).
\newblock Motion optimization for musculoskeletal dynamics: A flatness-based polynomial approach.
\newblock \emph{IEEE Trans. Automat. Contr.}, {\tt DOI: 10.1109/TAC.2020.3029318}

\bibitem{sadeghi}
Sadeghi M., Greene J.M., Sontag E.D. (2021).
\newblock Universal features of epidemic models under social distancing guidelines.
\newblock \emph{Annual Rev. Contr.}, 51, 426-440.

\bibitem{sahoo}
Sahoo S.R., Chiddarwar S.S. (2020).
\newblock Flatness-based control scheme for hardware-in-the-loop simulations of omnidirectional mobile robot.
\newblock \emph{Simul.}, 96, 169-183.

\bibitem{sancak}
Sancak C., Yamac F., Itik, M., Alici G. (2021).
\newblock Force control of electro-active polymer actuators using model-free intelligent control.
\newblock \emph{J. Intel. Mater. Syst. Struct.}, {\tt https://doi.org/10.1177/1045389X20986992}

\bibitem{sanchez}
Sanchez J.C., Gavilan F., Vazqueza R., Louembet C. (2020).
\newblock A flatness-based predictive controller for six-degrees of freedom spacecraft rendezvous.
\newblock \emph{Acta Astronaut.}, 167,  391-403.


\bibitem{schorghuber}
Schörghuber C., G\"{o}lles M., Reichhartinger M., Horn M. (2020).
\newblock Control of biomass grate boilers using internal model control.
\newblock \emph{Contr. Eng. Pract.}, 96, 104274.

\bibitem{sehili}
Sehili L., Boukhezzar B. (2022).
\newblock  Ultra-local model design based on real-time algebraic and derivative estimators for position control of a DC motor.
\newblock \emph{J. Contr. Autom. Electr. Syst.}, {\tt https://doi.org/10.1007/s40313-021-00881-z}

\bibitem{sekiguchi}
Sekiguchi K., Eikyu W., Nonaka K. (2021).
\newblock Feedback control for a drone with a suspended load via hierarchical linearization.
\newblock \emph{J. Robot. Mechatron.}, 33, 274-282.


\bibitem{sontag}
Sontag E.D. (2021)..
\newblock An explicit formula for minimizing the infected peak in an SIR epidemic model when using a fixed number of complete lockdowns.
\newblock \emph{Int. J. Robust Nonlin. Contr.}, {\tt https://doi.org/10.1002/rnc.5701}

\bibitem{srour}
Srour A., Noura H., Theilliol (2021).
\newblock  Passive fault-tolerant control of a fixed-wing UAV based on model-free control.
\newblock \emph{5th Conf. Contr. Fault Toler. Syst. (SysTol)}, Saint-Rapha\"{e}l.

\bibitem{steckler}
Steckler P.-B., Gauthier J.-Y., Lin-Shi X., Wallart (2021).
\newblock Differential flatness-based, full-order nonlinear control of a modular multilevel converter (MMC).
\newblock \emph{IEEE Trans. Contr. Syst. Techno.}, {\tt DOI: 10.1109/TCST.2021.3067887}

\bibitem{stella}
Stella L., Pinel Martínez A., Bauso D., Colaneri P. (2022).
\newblock The role of asymptomatic infections in the COVID-19 epidemic via complex networks and stability analysis.
\newblock \emph{SIAM J. Contr. Optim.}, S119-S144.


\bibitem{sun}
Sun J., Wang J., Yang P., Guo S. (2021).
\newblock Model-free prescribed performance fixed-time control for wearable exoskeletons.
\newblock \emph{Appl. Math. Model.}, 90, 61-77.


\bibitem{tal}
 Tal E.A., Karaman S. (2021).
\newblock Global trajectory-tracking control for a tailsitter flying wing in agile uncoordinated flight.
\newblock \emph{AIAA  Aviat. Forum}.


\bibitem{thounthong}
Thounthong P., Mungporn P., Guilbert D., Takorabet N., Pierfedericie S., Nahid-Mobarakehf B., Hug Y., Bizonh N., Huangfui Y., Kumamj P. (2021).
\newblock Design and control of multiphase interleaved boost converters-based on differential flatness theory for PEM fuel cell multi-stack applications.
\newblock \emph{Elect. Power Ener. Syst}, 124, 106346.

\bibitem{tognon}
Tognon M., Franchi A. (2021). 
\newblock \emph{Theory and Applications for Control of Aerial Robots in Physical Interaction Through Tethers}. 
\newblock Springer.

\bibitem{truong}
Truong C.T., Huynh K.H., Duong V.T., Nguyen H.H., Pham L.A., Nguyen T.T. (2021).
\newblock Model-free volume and pressure cycled control of automatic bag valve mask ventilator.
\newblock \emph{AIMS Bioengin.}, 8, 192-207.

\bibitem{tsay}
Tsay C., Lejarza F., Stadtherr M.A., Baldea M. (2020).
\newblock Modeling, state estimation, and optimal control for the US COVID-19 outbreak.
\newblock \emph{Scientif. Rep.}, 10, 10711.


\bibitem{xu0}
Xu L., Chen G., Li Q. (2020).
\newblock Ultra-local model-free predictive current control based on nonlinear disturbance compensation for permanent magnet synchronous motor.
\newblock \emph{IEEE Access}, 8,  127690-127699.

\bibitem{xu1}
Xu L., Chen G., Li Q. (2021).
\newblock Cascaded speed and current model of PMSM with ultra-local model-free predictive control.
\newblock \emph{Int. J. Robust Nonlinear Contr.}. {\tt \scriptsize https://doi.org/10.1049/elp2.12108}

\bibitem{villagra}
Villagra, J., Herrero-P\'{e}rez, D. (2012).
\newblock A comparison of control techniques for robust docking maneuvers of an AGV.
\newblock \emph{IEEE Trans. Contr. Syst. Techno.}, 20, 1116-1123.


\bibitem{spva} 
Wang Z., Bauch C.T., Bhattacharyya S., d'Onofrio A., Manfredi P., Perc M., Perra N., Salath\`{e} M., Zhao D. (2013). 
\newblock Statistical physics of vaccination. 
\newblock \emph{Phys. Rep.}, 664, 1-113.

\bibitem{wangc}
Wang Z., Cosio A., Wang J. (2022).
\newblock Implementation resource allocation for collision-avoidance assistance systems considering driver capabilities.
\newblock \emph{IEEE Trans. Intel. Transport. Syst.}

\bibitem{wanga}
Wang Y., Li H., Liu R., Yang L., Wang X. (2020a).
\newblock Modulated model-free predictive control with minimum switching losses for PMSM drive system.
\newblock \emph{IEEE Access}, 8, 20942-20953.

\bibitem{wangb}
Wang Z., Wang J. (2020b). 
\newblock Ultra-local model predictive control: A model-free approach and its application on automated vehicle trajectory tracking.
\newblock \emph{Contr. Eng. Pract.},  101, 104482.

\bibitem{weiss}
Weiss H. (2013).
\newblock The SIR model and the foundations of public health.
\newblock \emph{Materials matem\`{a}tics},  {\tt https://ddd.uab.cat/record/108432}

\bibitem{zauner}
Zauner M., Mandl P., K\"{o}nig O., Hametner C., Jakubek S. (2021).
\newblock Stability analysis of a flatness-based controller driving a battery emulator with constant power load.
\newblock \emph{at-Automatisierungstech}, 69,142-154.

\bibitem{zhang0}
Zhang Y., Jiang T., Jiao J. (2020).
\newblock Model-free predictive current control of a DFIG using an ultra-local model for grid synchronization and power regulation.
\newblock  \emph{IEEE Trans. Energ. Conv.}, 35, 2269-2280.

\bibitem{zhang1}
Zhang Y., Wang X., Yang H., Zhang B., Rodriguez J. (2021).
\newblock Robust predictive current control of induction motors based on linear extended state observer.
\newblock \emph{Chinese J. Elec. Engin.}, 7, 94-105.

\bibitem{zhou}
Zhou Z., Wang Z., Wang J. (2021).
\newblock  
Real-time adaptive threshold adjustment for lane detection application under
different lighting conditions using model-free control.
\newblock \emph{IFAC PapersOnLine}, 54-20, 147-152.


}

\end{thebibliography}
\end{document}